\documentclass[lettersize,journal]{IEEEtran}
\usepackage{amsmath,amsfonts}
\usepackage{algorithmic}
\usepackage{array}
\usepackage[caption=false,font=footnotesize,labelfont=rm,textfont=rm]{subfig}
\usepackage{textcomp}
\usepackage{stfloats}
\usepackage{url}
\usepackage{subfig}
\usepackage{verbatim}
\usepackage{graphicx}
\usepackage{bm}
\usepackage{amssymb}
\usepackage{booktabs}
\usepackage{cite}
\usepackage{color}
\usepackage{caption}
\def\BibTeX{{\rm B\kern-.05em{\sc i\kern-.025em b}\kern-.08em
    T\kern-.1667em\lower.7ex\hbox{E}\kern-.125emX}}
\usepackage{balance}
\begin{document}
\title{\huge{{Synesthesia of Machines (SoM)-Enhanced Wideband Multi-User CSI Learning With LiDAR Sensing}}}
\author{Haotian Zhang,~\IEEEmembership{Graduate Student Member,~IEEE,} Shijian Gao,~\IEEEmembership{Member,~IEEE,}  Xiang Cheng,~\IEEEmembership{Fellow,~IEEE,}  Liuqing Yang,~\IEEEmembership{Fellow,~IEEE} 

	\thanks{
    This work was supported in part by the National Natural Science Foundation of China under Grant 62125101, Grant 62341101, and Grant 62401488, and the New Cornerstone Science Foundation through the Xplorer Prize; in part by Guangzhou-HKUST(GZ) Joint Funding Scheme under No. 2025A03J3878, and the Guangdong Provincial Key Lab of Integrated Communication, Sensing and Computation for Ubiquitous Internet of Things under No. 2023B1212010007; in part by Natural Science Foundation of China Project \#U23A20339, Guangzhou Municipal Science and Technology Project \#2023A03J0011, and Guangdong Provincial Project \#2023ZDZX1037 and \#2023ZT10X009. (\textit{Corresponding Authors: Xiang Cheng; Shijian Gao.})
    
	Haotian Zhang and Xiang Cheng are with the State Key Laboratory of Photonics and Communications, School of Electronics, Peking University, Beijing 100871, P. R. China (e-mail: haotianzhang@stu.pku.edu.cn; xiangcheng@pku.edu.cn).

    Shijian Gao is with the Internet of Things Thrust, The Hong Kong University of Science and Technology (Guangzhou), Guangzhou 511400, P. R. China, and the Guangdong Provincial Key Laboratory of Future Networks of Intelligence, The Chinese University of Hong Kong, Shenzhen, P. R. China (e-mail: shijiangao@hkust-gz.edu.cn).

    Liuqing Yang is with the Internet of Things Thrust and Intelligent Transportation Thrust, The Hong Kong University of Science and Technology (Guangzhou), Guangzhou 510000, P. R. China, and the Department of Electronic and Computer Engineering, The Hong Kong University of Science and Technology, Hong Kong SAR, P. R. China (email: lqyang@ust.hk).

}}

\maketitle

\begin{abstract}
  Light detection and ranging (LiDAR) has been utilized for optimizing wireless communications due to its ability to detect the environment. This paper explores the use of LiDAR in channel estimation for wideband multi-user multiple-input-multiple-output orthogonal frequency division multiplexing systems and introduces a LiDAR-Enhanced Channel State Information (CSI) Learning Network (LE-CLN). By utilizing user positioning information, LE-CLN first calculates user-localized over-complete angular measurements. It then investigates the correlation between LiDAR and CSI, transforming raw LiDAR data into a low-complexity format embedded with signal propagation characteristics. LE-CLN also adapts the use of LiDAR based on channel conditions through attention mechanisms. Thanks to the unique wireless features offered by LiDAR, LE-CLN achieves higher estimation accuracy and spectrum efficiency compared to benchmarks, particularly in latency-sensitive applications where pilot transmissions are expected to be reduced.
  
\end{abstract}


\section{Introduction}

\IEEEPARstart{L}{ight} detection and ranging (LiDAR) has become an essential sensor in intelligent transportation systems owing to its precise and long-range environment detection capabilities. Nowadays, researchers have increasingly turned their attention to the use of LiDAR in wireless communications \cite{ klautau2019lidar, alkhateebLiDARBeam, LiDAR0, LiDAR2, LiDAR4}, as the high-frequency wireless channel and environmental geometric features are closely interconnected \cite{SoM,ISAC-VCN}. Multi-modal sensing that characterizes the environmental geometric features and communications can promote each other via the Synesthesia of Machines (SoM) \cite{SoM}. Previous studies \cite{klautau2019lidar, alkhateebLiDARBeam, LiDAR0, LiDAR2} have utilized LiDAR to aid in predicting optimal beam indexes or link blockage status, thereby reducing the costly overhead associated with beam training. Additionally, in another study~\cite{LiDAR4}, LiDAR was employed to improve user localization accuracy by extracting geometric information about scatterers.

Channel state information (CSI) is essential for the operation of communication systems. In millimeter-wave (mmWave) multiple-input multiple-output (MIMO) system, channel estimation faces thorny challenges due to the use of hybrid (analog/digital) structure and high-dimensional CSI. Conventional channel estimation schemes \cite{ wei2021beamspace,ma2020sparse,tongBCE,DSDS} often experience performance degradation in scenarios where limited pilot transmissions are available, as they rely solely on obtaining wireless channel features from pilots.
In response to this issue, Jiang \textit{et al}. \cite{sensingaidedCE2022} proposed to reduce the pilot overhead by offering initial estimated support of the sparse signal for orthogonal matching pursuit (OMP) by inferring the delay, Doppler, and angles of departure from radar signals. {\color{black}However, some peaks detected from the radar signals may be unrelated to the communication paths. Indiscriminately using prior information from all detected peaks could decrease channel estimation accuracy. Xu \textit{et al}. \cite{xu2021deep}} proposed to leverage scene images and other side information to realize pilot-free channel covariance matrix estimation. Nonetheless, the instant CSI estimation depends on the linear minimum mean square error algorithm and requires a certain number of pilots. The LiDAR-aided pilot-free beam prediction technique \cite{klautau2019lidar, alkhateebLiDARBeam, LiDAR0, LiDAR2} is capable of estimating only the approximate angle of the channel's line-of-sight (LoS) component. Obtaining such limited channel knowledge is insufficient for practical and complex channels. As a step forward, our previous work \cite{MMFF} introduces the intelligent fusion of RGB images, depth maps, and CSI at sub-6 GHz channels to achieve the accurate angle estimation of mmWave channel's LoS component.

Though the above works have attempted to leverage out-of-band (OOB) sensing information to reduce pilots, research on exploring the use of LiDAR in channel estimation for wideband multi-user MIMO orthogonal frequency division multiplexing (OFDM) system is currently lacking. To bridge this gap, this paper proposes a LiDAR-enhanced CSI learning network (LE-CLN) under the guidance of SoM \cite{SoM}, realizing the intelligent enhancement of communication functionality through multi-modal sensing. The effective and efficient utilization of LiDAR in LE-CLN is manifested in three aspects. Firstly, LE-CLN establishes a user-localized over-complete discrete Fourier transform (ULO-DFT) codebook to derive more angular measurements from limited pilot observations based on the user positioning information. Secondly, LE-CLN transforms raw LiDAR data into a low-complexity format and integrates prior information closely linked to CSI by incorporating the large-scale and small-scale fading characteristics into the LiDAR data.
Thirdly, LE-CLN allows for intelligent adjustment of the usage level of LiDAR according to the channel condition by a weight control mechanism. Finally, LE-CLN enables flexible wideband multi-user downlink transmission by a convolutional neural network (CNN) based channel interpolation module.

In conclusion, LE-CLN stands out as an innovative channel estimator tailored for wideband multi-user MIMO-OFDM systems, effectively utilizing the inherent correlation between LiDAR environmental detection results and CSI. Through simulations, LE-CLN demonstrates substantial enhancements in channel estimation accuracy by integrating out-of-band wireless channel insights obtained from LiDAR, particularly in scenarios with limited pilot signals.\footnote{Simulation codes are provided to reproduce the results presented in this paper: https://github.com/Haotian-Zhang-PKU/CSI-Learning-LECLN}

\section{System Model}
\begin{figure}[!t]
	\centering
	\includegraphics[width=1\linewidth]{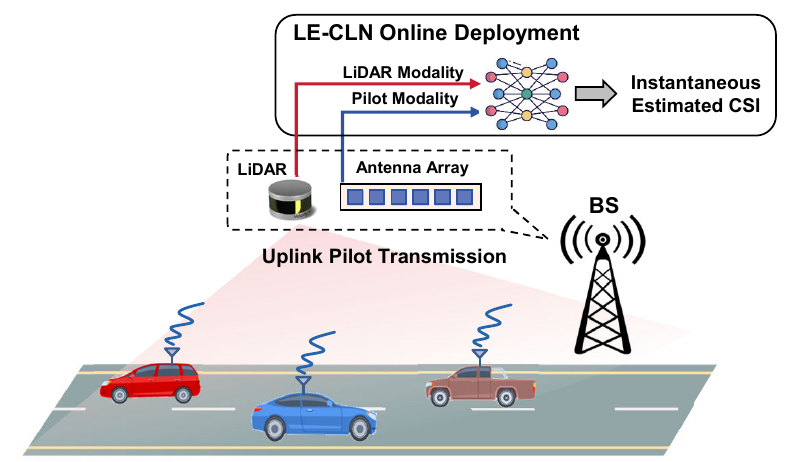}
	\caption{An illustration of the system model. 
	\label{system model}}
\end{figure}

{\color{black}We consider a time division duplex (TDD) based multi-user wideband hybrid MIMO-OFDM communication system where a BS equipped with a uniform linear array uses $N_{\rm{s}}$ subcarriers to serve $K$ single-antenna users, as shown in Fig. \ref{system model}.}  The BS is equipped with $N_{\rm{RF}}$ RF chains and $N_{\rm{t}}$ antennas (generally $N_{\rm{RF}} < N_{\rm{t}}$). The digital precoding matrix and analog precoding matrix adopted by BS are denoted by $\mathbf{F}_{\rm{B}} \in \mathbb{C}^{N_{\rm{RF}} \times N_{\rm{RF}}}$ and $\mathbf{F}_{\rm{R}} \in \mathbb{C}^{N_{\rm{t}} \times N_{\rm{RF}}}$ with $\Vert \mathbf{F}_{\rm{R}}\mathbf{F}_{\rm{B}} \Vert_{\rm{F}}^2=N_{\rm{RF}}$.  {\color{black}$\mathbf{F}_{\rm{B}}$ is set as an identity matrix $\mathbf{I}_{N_{\rm{RF}}}$ similar to \cite{ma2020sparse,digital3} and is omitted in the following contents.} {\color{black}Leveraging the channel reciprocity, the downlink channel can be directly derived through estimating the uplink channel. }

In the considered system, multiple users adopt orthogonal subcarriers to transmit pilots to BS. The BS then estimates the channels of these subcarriers, allowing for the recovery of full downlink channels. Let $\mathbb{V}_k = \{v_1,v_2,\cdots,v_{K_{\rm{P}}}\}$ represent the set of subcarrier indices used by the $k$-th user for pilot transmission, where the subcarrier interval $\Delta_v = v_{i+1}-v_i = \lfloor {N_{\rm{s}}}/{K_{\rm{P}}} \rfloor$. It follows $\mathbb{V}_i \cap \mathbb{V}_j = \varnothing$ for $i \neq j$. 

For the $k$-th user, the pilot symbols transmitted are denoted by $\mathbf{S}^{\rm{P}}_k \in \mathbb{C}^{K_{\rm{P}} \times N_{\rm{P}}}$ satisfying $\mathbb{E}\{\mathbf{S}^{\rm{P}}_k(\mathbf{S}^{\rm{P}}_k)^{\rm H}\}=\mathbf{I}_{K_{\rm{P}}}$, where $K_{\rm{P}}$ and $N_{\rm{P}}$ represent the number of pilots placed along the subcarrier and time-slot axes. The full downlink channel between the BS and the $k$-th user is denoted by $\mathbf{H}_k\triangleq[\mathbf{h}_k^1, \mathbf{h}_k^2, \cdots, \mathbf{h}_k^{N_{\rm{s}}}] \in \mathbb{C}^{ N_{\rm{t}} \times N_{\rm{s}}}$, where $\mathbf{h}_k^m \in \mathbb{C}^{N_{\rm{t}} \times 1}$ is the channel over the $m$-th subcarrier for $m=1,2,\cdots,N_{\rm{s}}$. After the pilot transmission, the overall measurements $\mathbf{Y}^{\rm{P}}_k \in \mathbb{C}^{N_{\rm{RF}}\times  N_{\rm{P}}}$ obtained by BS can be expressed as 
\begin{equation}
\label{m}
	\mathbf{Y}^{\rm{P}}_k   = \mathbf{F}_{\rm{R}}^{\rm H} \mathbf{H}^{\rm{P}}_k\mathbf{S}^{\rm{P}}_k + \mathbf{F}_{\rm{R}}^{\rm H} \mathbf{N}_k,
\end{equation}
where $\mathbf{H}^{\rm{P}}_k \in \mathbb{C}^{N_{\rm{t}} \times K_{\rm{P}}}$ represents the channel at pilot positions with $K_{\rm{P}}=\lvert \mathbb{V} \rvert $ and $\mathbf{N}_k \in \mathbb{C}^{N_{\rm{t}} \times N_{\rm{P}}}$ represents the additive white Gaussian noise with $n_{i,j} \sim \mathcal{CN}(0,\sigma^2)$. 

Similar to  \cite{Channel model}, the channel at the $m$-th subcarrier is modeled as: $\mathbf{h}_m = \sum_{l=1}^L \alpha_l  e^{j(\phi_l-2\pi\frac{ m}{N_{\rm{s}}}\tau_l)} \mathbf{a}(\theta_l)$
where $L$ is the number of channel paths and $T_{\rm{s}}$ is the sampling interval. {\color{black}For the $l$-th path}, $\theta_l$ is the azimuth AoA, $\tau_l = {\tilde{\tau_l}}/{T_{\rm{s}}}$ is the normalized path delay, $\tilde{\tau_l}$ is the path delay,  $\phi_l$ is the phase shift of the $l$-th path, {\color{black}and $\mathbf{a}(\theta_l) \in \mathbb{C}^{N_{\rm{t}}\times 1}$ is the array response vector at $ \theta_l$ with $[\mathbf{a}(\theta_l)]_i = \frac{1}{\sqrt{N_{\rm{t}}}}e^{-{\rm{j}}\pi(i-1)\theta_l}$.}

\begin{figure*}[!t]
	\centering
        \captionsetup{justification=centering}
	\includegraphics[width=1\linewidth]{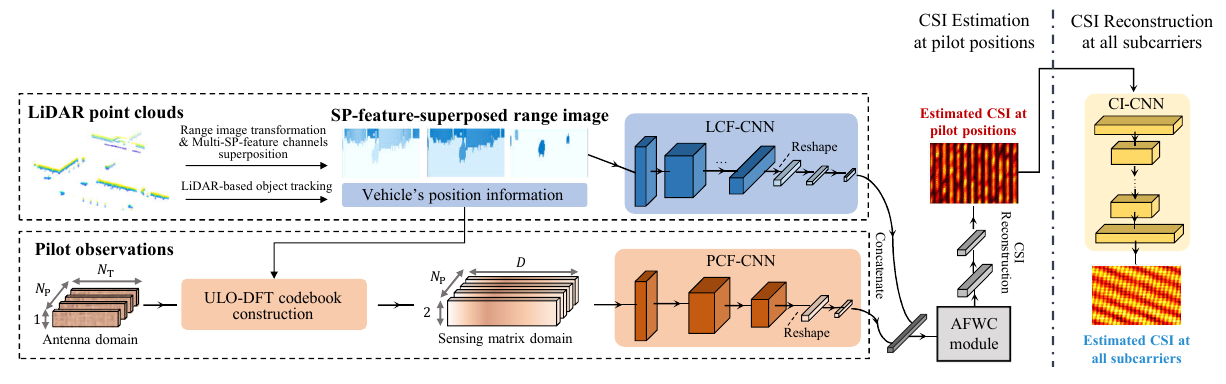}
	\caption{{\color{black}{The schematic of proposed LE-CLN framework.}}}
	\label{scheme}
\end{figure*}

\section{Proposed LiDAR-Enhanced CSI Learning Network (LE-CLN)}
In this section, we start with the preliminary information obtained from LiDAR. We then proceed to discuss the technical details of the proposed LE-CLN. The overall architecture of the LE-CLN model is shown in Fig. \ref{scheme}.
\subsection{Coarse LiDAR Processing}
Assume that the BS has prior knowledge of road positions. Also, the BS can access the vehicle's initial location through the Global Positioning System and track it using a sophisticated LiDAR-based object tracking algorithm. The process involves the following steps:

\textit{\textbf{Step 1: Raw Data Filtering}}. We first identify the ground points and eliminate them from the entire point cloud though they may contain multipath information from ground reflections. This is because the NN can hardly acquire such information due to its complex generation process. The data filtering procedure will alleviate the computational burden while retaining the primary useful features. It is worth noting that in cases where multipath components from ground reflections are non-negligible, more pilots will be needed, to supply features that cannot be acquired through LiDAR.

\textit{\textbf{Step 2: Density-Based Spatial Clustering of Applications with Noise (DBSCAN) Clustering}}. DBSCAN is utilized to segment the point cloud into different clusters. Then, BS can distinguish the receiving vehicle therein and assign the corresponding points to category $\mathcal{V}_{\rm{r}}$. It can also categorize the points corresponding to the vehicles that are close to the receiving vehicle as $\mathcal{V}_{\rm{s}}$. The angle of the receiving vehicle relative to BS $\theta_k$ can be roughly determined based on the centroid of the corresponding cluster.

\subsection{\color{black}{CSI Feature Extraction via Coarse LiDAR Processing}}
\label{SE}
In the context of limited pilot transmissions, we propose constructing a ULO-DFT codebook $\mathbf{A} \in \mathbb{C}^{N_{\rm{t}} \times D}$ to enable LE-CLN to obtain more fine-grained angular measurements near the user, identifying possible locations of dominant entries. Unlike a typical DFT codebook whose steering angles are uniformly distributed over $[0, 2\pi)$, {\color{black}the region close to user is divided into multiple angle grids. The entire angular space is partitioned into an oversampled angular subspace along with the non-oversampled ones. Hence, the size of the ULO-DFT codebook could be larger than the number of antennas, i.e., $D \gg N_{\rm t}$. }
Let $[\phi_{\rm{o}_1}, \phi_{\rm{o}_2})$ represent the over-sampled angular subspace. The endpoints are determined as follows: 
\begin{subequations}
\begin{equation}
\phi_{\rm{o}_1}=
\left\{
    \begin{array}{cl}
    \epsilon\lfloor\frac{\theta_k-\epsilon\frac{(N_{\rm{w}}-1)}{2}}{\epsilon} \rfloor&,  ~\theta_k-\epsilon\frac{(N_{\rm{w}}-1)}{2} \geq 0  \\
    0&,~~\text{otherwise}
     \end{array}
 \right.  
\end{equation}
\begin{equation}
 \phi_{\rm{o}_2}=\epsilon\lfloor\frac{\theta_k+\epsilon\frac{(N_{\rm{w}}-1)}{2}}{\epsilon} 
 \rfloor,
 \end{equation}
\end{subequations}
where $N_{\rm{w}}$ determines the size of over-sampled subspace; $\epsilon={2\pi}/{2N_{\rm{t}}}$ is the sampling interval of a uniform over-complete DFT codebook. 
The overall angle grid distribution $\mathbf{\Phi}$ is given by:
\begin{equation}
\label{phi}
  \mathbf{\Phi} = \left\{
		\begin{array}{cl}
           \bigl[0,\cdots,\phi_{\rm{o}_2}, \Big| \phi_{\rm{g}_1},\cdots,2\pi 
         \bigr], ~\phi_{\rm{o}_1} = 0& \\
         \bigl[
         0,\cdots,\phi_{\rm{g}_1}, \Big| \phi_{\rm{o}_1},\cdots,\phi_{\rm{o}_2}, \Big| \phi_{\rm{g}_2},\cdots,2\pi
         \bigr], ~\text{otherwise}&
         \end{array} 
  \right.
\end{equation}
where $\phi_{\rm{g}_1}$ and $\phi_{\rm{g}_2}$ are the end points of the plain angle subspace. Note that the angle grid within the over-sampled and plain subspace is uniformly partitioned, with an angle resolution of $\epsilon_{\rm{o}}={\epsilon}/{2}$ in the over-complete subspace and $\epsilon_{\rm{g}}=({2\pi-N_{\rm{o}}\epsilon_{\rm{o}}})/({D-N_{\rm{o}}})$ in the plain subspace. $N_{\rm{o}}$ is the number of angle grids in the over-sampled subspace, which is determined by $N_{\rm{w}}$. For instance, considering the second distribution case in Eq. \eqref{phi} as a reference, the angle grid $\phi_\emph{i}$ in different subspaces can be calculated as:
\begin{equation}
  \phi_\emph{i} =
  \left\{
		\begin{array}{cl}
			 (i-1)\epsilon_{\rm{g}},~\emph{i} \in [1,{\rm{o}_1})& \\
                \phi_{\rm{o}_1} + (i-\rm{o}_1)\epsilon_{\rm{o}},~\emph{i} \in [{\rm{o}_1},{\rm{o}_2}]& \\
			\phi_{\rm{o}_2} + (i-\rm{o}_2)\epsilon_{\rm{g}},~ \emph{i} \in (\rm{o}_2,\emph{D}]&.
		\end{array} 
  \right.
\end{equation}
Finally, the ULO-DFT codebook $\mathbf{A}$ is constructed as $[\mathbf{a}(\phi_1),\cdots,\mathbf{a}(\phi_D)]$.

Based on $\mathbf{A}$, the antenna-domain channel can be transformed into angular domain by $\mathbf{H}^{\rm{a}}_k = \mathbf{A}^{\rm H}\mathbf{H}_k$, where $\mathbf{H}^{\rm{a}}_k \in \mathbb{C}^{D\times K_{\rm{P}}}$ is the angular-domain channel between the $k$-th user and BS of the pilot position. 
Using $(\mathbf{A}^{\rm H})^\dagger\mathbf{A}^{\rm H}=\mathbf{I}_{N_{\rm t}}$, the received pilot signals can be rewritten as\footnote{ $\mathbf{A}$ is a row full rank matrix. $(\mathbf{A}^{\rm H})^\dagger = \mathbf{A}(\mathbf{A}^{\rm H}\mathbf{A})^{-1}$.}: 
 \begin{equation} 
    \mathbf{Y}^{\rm{P}}_k 
     = \underbrace{\mathbf{F}_{\rm{R}}^{\rm H} (\mathbf{A}^{\rm H})^{\dagger}}_{\Theta}\mathbf{H}^{\rm{a}}_k\mathbf{S}^{\rm{P}}_k +  \mathbf{F}_{\rm{R}}^{\rm H} \mathbf{N}^{\rm{P}}_k.  \label{eq:eq2}
\end{equation}
Here, $\Theta= \mathbf{F}_{\rm{R}}^{\rm H} (\mathbf{A}^{\rm H})^{\dagger}$ is defined as the sensing matrix. Then, the pilot observations $\mathbf{Y}^{\rm{P}}_k$ are projected into the sensing matrix space by $\tilde{\mathbf{Y}}^{\rm{P}}_k=\Theta^{\dagger}\mathbf{Y}^{\rm{P}}_k$, which captures user-localized angular features. The complex value $\tilde{\mathbf{Y}}^{\rm{P}}_k \in \mathbb{C}^{D\times N_{\rm{p}}}$ is then transformed into real value input $\tilde{\mathbf{Y}}^{\rm{r}}_k \in \mathbb{R}^{2\times D \times N_{\rm{p}}}$ by treating the imaginary and real parts as two parallel dimensions.
$\tilde{\mathbf{Y}}^{\rm{r}}_k$ serves as the input for a customized two-layer CNN, referred to as the pilot-based CSI feature extraction CNN (PCF-CNN).

\subsection{OOB CSI Feature Extraction via Fine LiDAR Processing} 
In the situation where pilots offer only restricted wireless channel information, the objective is to uncover the inherent correlation between the environmental detection outcomes obtained from LiDAR and CSI. The goal is to transform the environmental detection findings from LiDAR into valuable out-of-band wireless channel characteristics.
In order to achieve this, a LiDAR processing technique is carefully devised. 
Note that, direct utilization of raw LiDAR data is avoided due to the substantial computational burden it imposes and the potential noise interference it may introduce to channel estimation. 

\textit{\textbf{Step 1: Lightweight Raw LiDAR Data Processing}}.
The raw LiDAR data $\mathcal{P}$ is initially transformed into the range image form $\mathbf{R}_{\rm{D}}$: $\mathbb{R}^3 \mapsto \mathbb{R}^2$. This conversion involves mapping each point  $\mathbf{p}_{i}=(x_i,y_i,z_i)$ to image coordinates $\mathbf{r}_i=(u_i, v_i)$ via Eq. \eqref{RD}. The pixel value in the range image corresponds to the range $r_i$ of each point, given by $r_i =\Vert \mathbf{p}_i \Vert_{\rm{2}}$. 
\begin{align}
	\begin{pmatrix} u_i \\ v_i \end{pmatrix} = \begin{pmatrix} \frac{1}{2}[1-\arctan(y_i,x_i) \cdot \pi^{-1}] \cdot w \\ [1-(\arcsin{(z_i,r_i^{-1})}+f_{\rm{up}}) \cdot f^{-1}]\cdot h \end{pmatrix}.
 \label{RD}
\end{align}
In Eq. \eqref{RD}, $f = f_{\rm{up}} + f_{\rm{down}}$ is the vertical field of view (FoV) of LiDAR; $f_{\rm{up}}$ and $f_{\rm{down}}$ are the upward and downward FoV, respectively. The width ($w$) and height ($h$) of range image are determined by the horizontal angular resolution $a_{\rm{h}}$ and the number of beams $M_{\rm{b}}$ of the LiDAR, such that $w = {360^{\circ}}/{a_{\rm{h}}}$ and $h = M_{\rm{b}}$. 

\textit{\textbf{Step 2: Multiple Signal Propagation (SP)-Feature Channels Superposition}}. 
Based on the LiDAR range image $\mathbf{R}_{\rm{D}}$, we propose to construct a multi-SP-feature superposed range image $\mathbf{R}_{\rm{D}}^{\rm{RF}} \in \mathbb{R}^{3 \times w \times h}$. The first channel represents the original range image, denoted as $\mathbf{R}_{\rm{D}}^{\rm{RF}}[1,u_i,v_i]=r_i$ where $r_i$ is the range at the pixel $\mathbf{r}_i$. The second channel is referred to as the equivalent small-scale fading map, which aims to provide information about potential multipath components. The pixel values in this channel, $\mathbf{R}_{\rm{D}}^{\rm{RF}}[2,u_i,v_i]$, can take one of three possible values: receiver, vehicles near the receiver (potential scatterers), and other objects. The third channel is known as the equivalent large-scale fading map, designed to offer insights into signal amplitude variation characteristics. The pixel values in this channel, $\mathbf{R}_{\rm{D}}^{\rm{RF}}[3,u_i,v_i]$, represent the path loss of the signal transmitted from transmitter to the point $\mathbf{r}_i$.  The pixel values of the second and third channels are determined through specific criteria and considerations related to the environmental detection results acquired from LiDAR, and are expressed as:
\begin{subequations}
\begin{equation} 
\mathbf{R}_{\rm{D}}^{\rm{RF}}[2,u_i,v_i]=
\left\{
    \begin{array}{cl}
    2,  ~&\mathbf{p}_{i} \in \mathcal{V}_{\rm{r}}  \\
    1,  ~&\mathbf{p}_{i} \in \mathcal{V}_{\rm{s}}  \\
    0,~&\text{otherwise}
     \end{array}
 \right.  
\end{equation}
\begin{equation}
    \mathbf{R}_{\rm{D}}^{\rm{RF}}[3,u_i,v_i]=40\lg \mathbf{R}_{\rm{D}}^{\rm{RF}}[1,u_i,v_i] + 20\lg f_{\rm{c}} - 20 \lg (h_{\rm{t}}h_{\rm{r}})
 \end{equation}
\end{subequations}
where $f_{\rm{c}}$ is the carrier frequency; $h_{\rm{t}}$ and $h_{\rm{r}}$ are the heights of transmitter and receiver. 

\textit{\textbf{Step 3: OOB CSI Feature Extraction}}. 
The constructed multi-SP-feature superposed range image $\mathbf{R}_{\rm{D}}^{\rm{RF}}$ is then used as the input for a customized five-layer CNN. This CNN is specifically designed for feature extraction and is referred to as the LiDAR-based CSI feature extraction CNN (LCF-CNN). The LCF-CNN processes the input data from the range image to extract relevant features that can be utilized to provide OOB wireless features.


\subsection{Feature Fusion via Attention}
Although OOB wireless features closely related to CSI have been extracted from LiDAR and fine-grained user-localized CSI features have been extracted from pilot observations, a key challenge that remains to be addressed is how to intelligently regulate the utilization of both modalities under varying channel conditions. Drawing inspiration from the SENet channel attention mechanism \cite{SENet}, we have designed an adaptive feature weight control (AFWC) module to facilitate element-wise weight allocation between CSI features from the two modalities. Denoting the CSI features extracted from pilots and LiDAR as $\mathbf{q}_{\rm{P}}$ and $\mathbf{q}_{\rm{L}}$, respectively, they are concatenated and input into the AFWC module,  which learns the importance of different features and outputs the weight vector $\mathbf{w}$:
\begin{equation}
\mathbf{w}  = \varphi_{\rm{w}}^3(\varphi_{\rm{w}}^2(\varphi_{\rm{w}}^1(\mathbf{q_{\rm P}}\oplus\mathbf{q_{\rm L}}))).
\end{equation}
Here, $\varphi_{\rm w}^l(\cdot) = f_{\rm{ReLU}}(\mathbf{W}_{\rm w}^{l}\mathbf{X} + \mathbf{b}_{\rm w}^{l}), l=1,2$ is the non-linear functions of the first and second fully connected (FC) layers with ReLU as the activation function; $\varphi_{\rm{w}}^3(\cdot)= f_{\rm{Sigmoid}}(\mathbf{W}_{\rm w}^{3}\mathbf{X} + \mathbf{b}_{\rm w}^{3})$ is the non-linear function of the third layer with Sigmoid as the activation function; $\oplus$ represents the concatenation operation.  $\mathbf{X}$ denotes a certain input, while $\mathbf{W}_{\rm w}^{l}$ and $\mathbf{b}_{\rm w}^{l}$ denote the weight and bias of the $l$-th layer, respectively. Next, a weighted feature $\mathbf{q}^{\rm{w}}$ is formed by performing an element-wise product between the weight vector $\mathbf{w}$ and the raw multi-modal CSI feature $\mathbf{q}=\mathbf{q_{\rm P}}\oplus\mathbf{q_{\rm L}}$, denoted as $\mathbf{q}^{\rm{w}} = \mathbf{q} \otimes \mathbf{w}$.

\subsection{CSI Reconstruction}
Using $\mathbf{q}^{\rm{w}}$ as input, and defining $\varphi_{\rm P}^{l}(\cdot)$ as the non-linear function of the $l$-th FC layer and $L_{\rm P}$ as the number of FC layers, a two-layer multilayer perceptron (MLP) $\text{MLP}_{\rm P}$ is designed to reconstruct the antenna-domain channel at all the pilot positions $\hat{\mathbf{H}}_k^{\rm{P}}$: 
\begin{equation}
    	\hat{\mathbf{H}}_k^{\rm{P}} = \text{MLP}_{\rm P}(\mathbf{q}^{\rm{w}}) = \varphi_{\rm P}^{L_{\rm P}}(
    	\varphi_{\rm P}^{L_{\rm P}-1}(\cdots
    	\varphi_{\rm P}^{1}(\mathbf{q}^{\rm{w}})\cdots)).
\end{equation}

Once CSI estimates at all pilot positions are obtained, a coarse CSI, denoted as ${\mathbf{V}}_k^{\rm{r}} \in \mathbb{C}^{N_{\rm t}\times N_{\rm s}}$ is acquired by padding the values at data positions with zeros. Then, a six-layer CNN is designed to recover the full CSI, termed as CSI interpolation CNN (CI-CNN). Its input $\hat{\mathbf{H}}_k^{\rm{r}} \in \mathbb{R}^{3 \times N_{\rm t}\times N_{\rm s}}$ is constructed by regarding the real part, imaginary part, and the phase of ${\mathbf{V}}_k^{\rm{r}}$ as three parallel channels. Then, the full antenna-domain channel $\hat{\mathbf{H}}_k$ is recovered by:
\begin{equation}
 \hat{\mathbf{H}}_k = \varphi_{\rm C}^{6}(\varphi_{\rm C}^{5}(\cdots \varphi_{\rm C}^{1}(\hat{\mathbf{H}}_k^{\rm{r}}) \cdots )).
\end{equation}
In this equation, $\varphi_{\rm C}^{l}(\mathbf{X}) = f_{\rm{ReLU}}(\mathbf{W}_{\rm C}^{l} \ast \mathbf{X} + \mathbf{b}_{\rm C}^{l})$ represents the function of the $l$-th convolutional layer,  `$\ast$' denotes the convolution operation, $\mathbf{W}_{\rm C}^{l}$ stands for the kernal, and $\mathbf{b}_{\rm C}^{l}$ indicates the bias.

{\color{black} Note that the CI-CNN module is trained independently. This separation is necessary due to the distinct mapping relationships learned by CI-CNN and the NN modules used for CSI recovery at pilot positions. The CI-CNN module models the relationships across different subcarriers, essentially serving the purpose of CSI interpolation in the frequency domain. The other NN modules in LE-CLN learn the mapping relationship between multi-modal input data and CSI at pilot positions. End-to-end training may make LE-CLN difficult to model these two distinct mappings, which may impact the ultimate estimation accuracy.}

\section{Simulation Results}
In this section, we present simulation results to evaluate the performance of the proposed LE-CLN.
In our simulations, both the wireless channel data and LiDAR data are derived from the vehicular urban crossroad scenario in the M$^3$SC dataset \cite{M3C}. The BS is equipped with $N_{\rm{t}}=32$ antennas. The carrier frequency is $f_{\rm{c}}=28$GHz and the bandwidth is $100$MHz consisting of $N_{\rm{s}} = 64$ subcarriers. We set the number of multiple paths as $n_{\rm{p}}=8$ and the heights of transceivers as $h_{\rm{t}}=5$m and $h_{\rm{r}}=1$m. The  size of the over-sampled subspace $N_{\rm{w}}$ is set to $13$. The parameters of LiDAR are: $a_{\rm{h}}=0.36^{\circ}$, $f_{\rm{up}}=15^{{\circ}}$, $f_{\rm{down}}=-25^{{\circ}}$, $M_{\rm{b}}=64$.
The resolution of the ULO-DFT codebook is set to $D=2N_{\rm{t}}=64$. The quantization bit number of the phase shifters adopted at BS is set to $B=3$, resulting in the quantization of the RF phase into $Q\triangleq2^B$ discrete values. Each element of $\mathbf{F}_{\rm{R}}$ is randomly selected from the set $\{{e^{\frac{j2\pi n}{Q}}, n = 1, 2,\cdots, Q}\}$. Table \ref{DL setup2} lists the hyper-parameters used for fine-tuning the LE-CLN model. All simulation results presented are averaged over $300$ independent realizations.
\begin{table}[!htp]
	\setlength{\abovecaptionskip}{0.1cm} 
	\renewcommand\arraystretch{0.05} 
	\centering
	\caption{Hyper-parameters for network fine-tuning}
	\label{DL setup2}
	\begin{tabular}{c|c}
		\toprule[0.35mm]
		\textbf{Parameter}	&\textbf{Value}  \\
		\midrule[0.15mm]
		Batch size	& 32 \\ 
		\midrule[0.15mm]
		Starting learning rate & $1 \times 10^{-3}$ \\ 
		\midrule[0.15mm]			
		Learning rate scheduler	& Epochs 80, 120, 150, 180 \\ 	
		\midrule[0.15mm]			
		Learning-rate decaying factor & 0.3 \\
		\midrule[0.15mm]		
		Epochs  & 300 \\ 	
		\midrule[0.15mm]		
		Optimizer  & ADAM \\ 
		\midrule[0.15mm]			
		Loss function  & MSEloss \\ 	
		\bottomrule[0.35mm]		
	\end{tabular}	
\end{table}

The proposed LE-CLN scheme is compared with the existing OMP~\cite{OMP}, AMP~\cite{AMP}, LS~\cite{LS}, CENN~\cite{ma2020sparse}, and GM-LAMP~\cite{wei2021beamspace} schemes.
The normalized mean square error (NMSE) is used as the estimation performance metric given by ${\rm NMSE}   =	{{\Vert \hat{\mathbf{H}}-\mathbf{H} \Vert}_{\rm{F}}^2}/{{\Vert \mathbf{H} \Vert}_{\rm{F}}^2}$. The number of measurements consumed by each scheme in the time domain is indicated in the legend. In the frequency domain, $8$ pilots are used and uniformly distributed across all subcarriers in all schemes. The average received SNR is defined as ${\rm SNR} = {\Vert \mathbf{F}_{\rm{R}}\mathbf{H}_k\Vert_{\rm{F}}^2}/({\sigma^2\Vert \mathbf{F}_{\rm{R}}\Vert_{\rm{F}}^2})$.

\begin{figure*}[!t]
	\centering\
         \captionsetup{justification=centering}
         \subfloat[]{\includegraphics[width=0.33\linewidth]{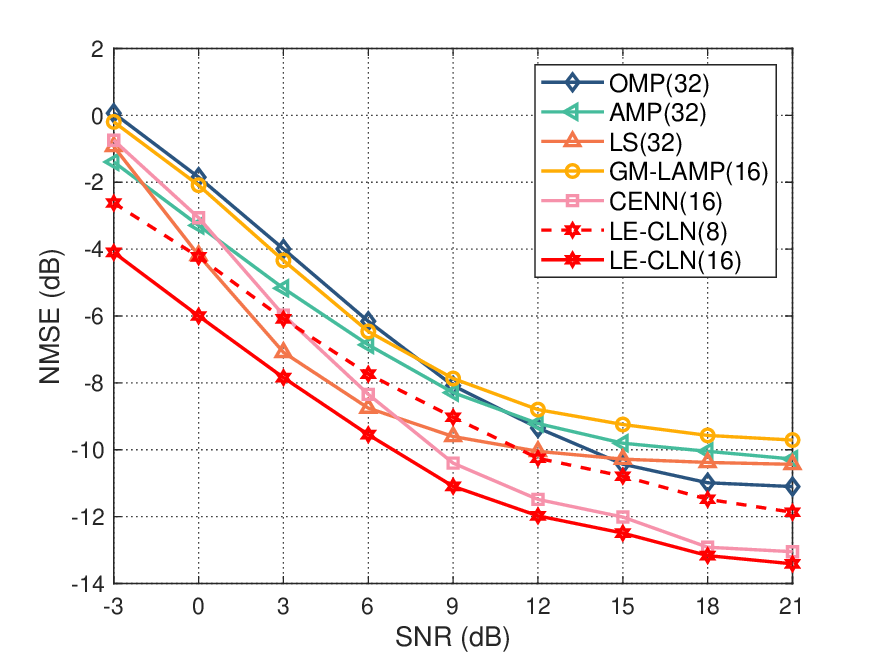}
         \label{nmse1}}
         \hspace {-5mm}
         \subfloat[]{\includegraphics[width=0.33\linewidth]{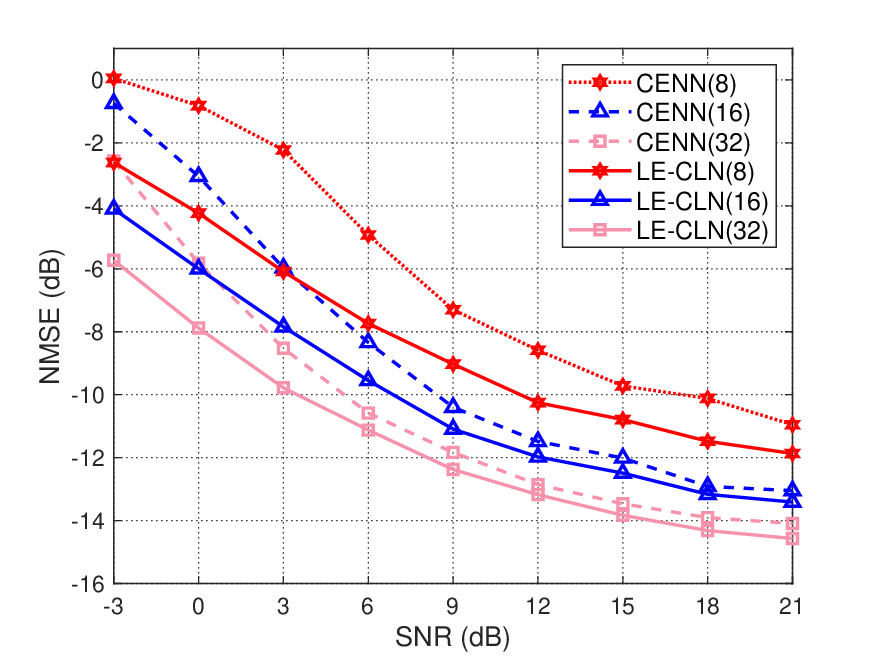}
         \label{nmse2}}
         \hspace {-5mm}
        \subfloat[]{\includegraphics[width=0.33\linewidth]{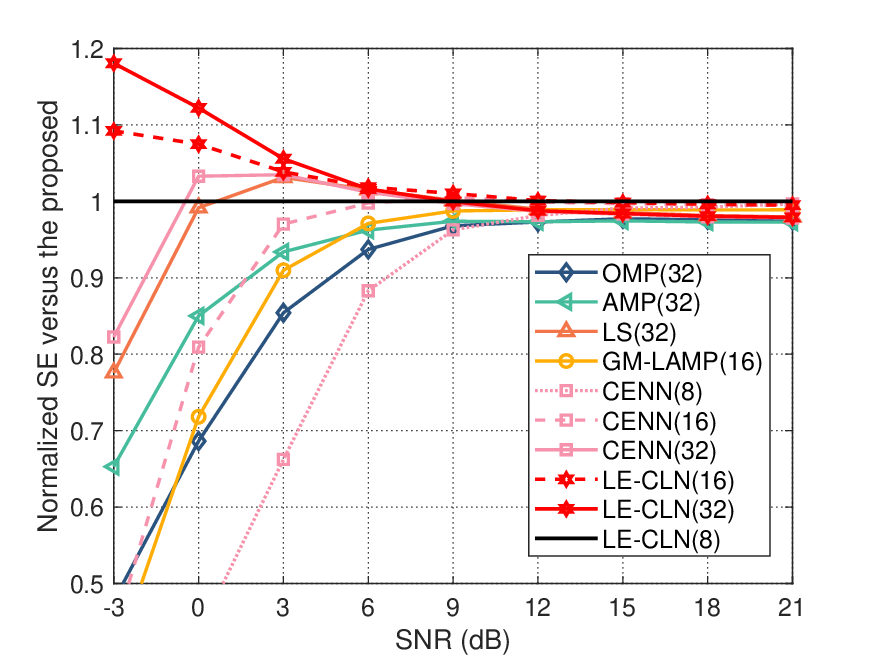}
         \label{se}}
	\caption{(a) NMSE performance comparisons for different schemes. (b) NMSE performance comparisons for CENN and LE-CLN with different number of measurements. (c) Normalized SE performance comparisons for different schemes versus the proposed LE-CLN(8) scheme.}
	\label{result2}
\end{figure*}

In Fig.~\ref{nmse1}, we first compare the NMSE performance of LE-CLN against the other schemes. LE-CLN achieves noticeable performance gain compared with OMP, AMP, and GM-LAMP schemes when the number of measurements used is the same. Furthermore, when the number of measurements used by LE-CLN is only $8$, its performance can be comparable to or even exceed those of OMP, AMP, and GM-LAMP with more measurements. As can be seen in Fig.~\ref{nmse2}, the performance improvement achieved by LE-CLN compared to CENN becomes significant as SNR and the available measurements decrease. This  improvement is attributed to the OOB wireless channel features provided by LiDAR. 

In Fig.~\ref{se}, we compare the upper bound of the downlink spectral efficiency (SE) of different schemes, which is measured by the fully-digital precoding. The zero-forcing (ZF) precoder at the $m$-th subcarrier can be calculated by $\mathbf{f}_k^{m}=[(\hat{\mathbf{h}}_k^m)^{\rm H} \hat{\mathbf{h}}_k^m]^{-1}(\hat{\mathbf{h}}_k^m)^{\rm H} $. The power of $\mathbf{f}_k^{m}$ is normalized by $\Vert \mathbf{f}_k^{m} \Vert_{\rm{2}}=1$. According to the specifications on the fifth-generation New Radio sidelink \cite{5GNR} and the channel coherence time in vehicular networks, we assume that total number of symbols is $900$, consisting of $N_{\rm{P}}$ pilot symbols and $N_{\rm{D}}$ data symbols. Then, the SE is given by $R_k = \frac{N_{\rm{D}}}{900} \frac{1}{N_{\rm{s}}}\sum_{m=1}^{N_{\rm{s}}}\log_{2}{(1+{|\mathbf{f}_k^{m}\mathbf{h}_k^m|^2}/{\sigma^2})}$ in the considered OFDM system. {\color{black}To clearly demonstrate the SE performance differences, Fig.~\ref{se} shows the normalized SE with the proposed LE-CLN(8)’s SE being the basis. The normalized SE is defined by the ratio between SE achieved by all schemes and that achieved by the proposed LE-CLN(8) scheme. A ratio greater than $1$ indicates that the SE achieved by a certain scheme is higher than that achieved by LE-CLN(8).} As can be seen, LE-CLN(8) achieves the highest SE in high SNRs ($[12,21]$dB). At low SNRs ($[-3,12]$dB), the LE-CLN(16), LE-CLN(32), LS(32), and CENN(32) schemes achieve higher SE performance owing to more accurate channel estimation results. Among these, the LE-CLN(32) scheme demonstrates the most optimal performance.

\begin{figure}[!t]
	\centering
    \captionsetup{justification=centering}
        \subfloat[]{\includegraphics[width=0.52\linewidth]{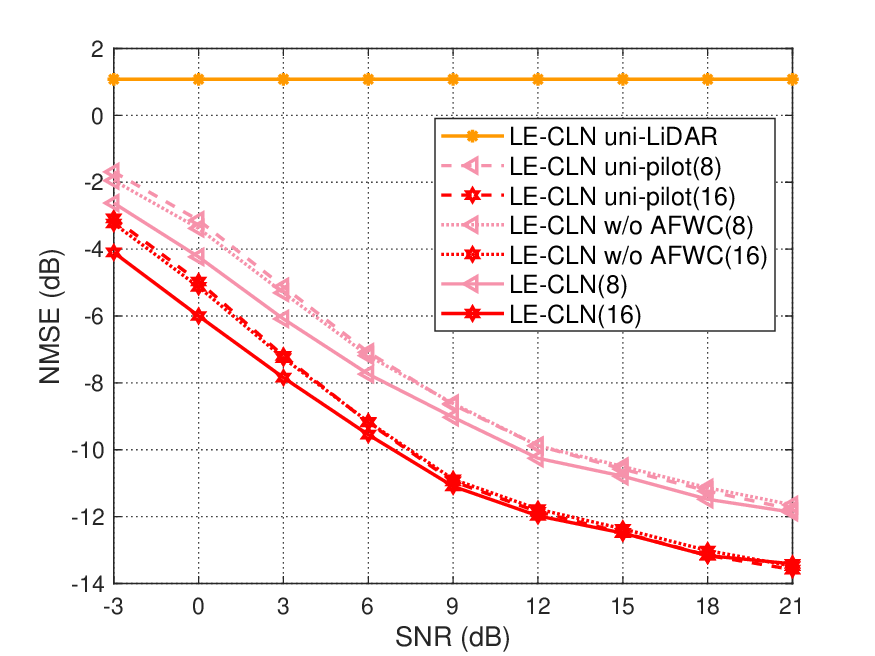}%
			\label{inner}}
   \hfill
   \hspace {-5mm}
        \subfloat[]{\includegraphics[width=0.52\linewidth]{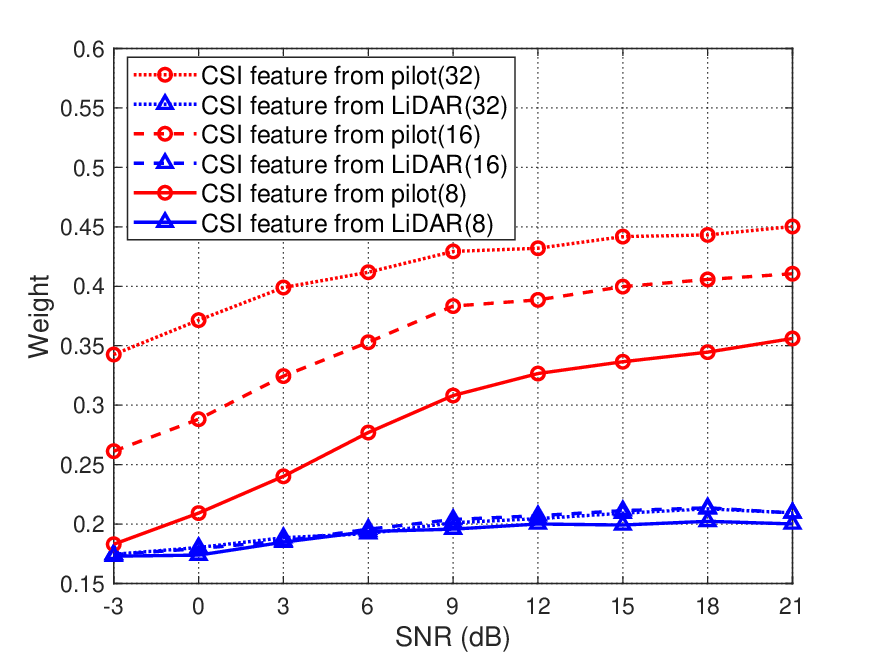}%
			\label{weight}}
	\caption{(a) NMSE performances of LE-CLN models with the absence of a certain module; (b) Attention weights allocated to multi-modal CSI features.}
	\label{result3}
 
\end{figure}
In Fig.~\ref{result3}, we demonstrate the roles of key modules in LE-CLN by removing them and evaluating the model's NMSE performance. The experimental results show that the absence of any key module leads to a performance degradation, and the extent of degradation increases as the measurements and SNR decrease.
{\color{black}It can be seen that only adopting LiDAR can achieve a rough channel estimation result, confirming that the proposed LiDAR utilization approach can provide effective OOB CSI features. Specifically, the multi-SP-feature LiDAR range image provides the approximate angles of the LoS and potential multipath components, along with large-scale fading information for these components. These basic information jointly contributes to the recovery of a rough CSI through NNs.} Furthermore, we can see that the performance of uni-pilot scheme gradually approaches that of LE-CLN, indicating that the contribution of OOB CSI features from LiDAR diminishes with higher SNRs. This aligns with the fact that pilots offer more intuitive and accurate CSI features at high SNRs. The performance gap between the LE-CLN without AFWC scheme and LE-CLN narrows as SNR increases for the same reason.  Additionally, in Fig.~\ref{weight}, the weight assigned to the pilot-originated CSI feature by the AFWC module noticeably increases with SNR. When the SNR is fixed, increasing the number of pilots results in a notable increase in the weights assigned to the pilot-originated CSI feature. These confirm that the AFWC module achieves an intelligent adjustment of the usage level of the multi-modal data under different channel conditions.

\section{Conclusion}
In this study, we investigated the use of LiDAR to improve channel estimation accuracy in wideband multi-user MIMO-OFDM systems. The LE-CLN model extracts additional angular features from pilots using the ULO-DFT codebook, which is constructed based on user positioning information. Furthermore, by integrating OOB wireless channel characteristics obtained from LiDAR environmental detection, LE-CLN achieves comparable or even superior estimation accuracy compared to standard benchmarks while reducing the number of required pilots. This performance enhancement is particularly pronounced in scenarios with lower SNRs and limited pilot measurements. Additionally, simulation results demonstrate that LE-CLN can dynamically adjust LiDAR utilization based on varying channel conditions, further enhancing its overall performance.


%

\end{document}